\documentclass{aa}  

\usepackage{graphicx}
\usepackage{txfonts}
\begin{document}

   \title{Suprathermal electron distributions in the solar transition region}

%   \subtitle{I. Overviewing the $\kappa$-mechanism}

   \author{C. Vocks\inst{1}
          \and
          E. Dzif\v{c}\'{a}kov\'{a},\inst{2}
          \and
          G. Mann\inst{1}
          }

   \institute{Leibniz-Institut f\"ur Astrophysik Potsdam,
              An der Sternwarte 16, 14482 Potsdam, Germany\\
              \email{cvocks@aip.de}
         \and
              Astronomical Institute of the Academy of Sciences of the
              Czech Republic, Fricova 298, 251 65 Ondrejov, Czech Republic\\
   }

   \date{Received; accepted}

% \abstract{}{}{}{}{} 
% 5 {} token are mandatory
 
  \abstract
  % context heading (optional)
  % {} leave it empty if necessary  
   {Suprathermal tails are a common feature of solar wind electron velocity
     distributions, and are expected in the solar corona. From the corona,
     suprathermal electrons can propagate through the
   steep temperature gradient of the transition region towards the
   chromosphere, and lead to non-Maxwellian electron velocity distribution
   functions (VDFs) with pronounced suprathermal tails.}
  % aims heading (mandatory)
   {We calculate the evolution of a coronal electron distribution through the
     transition region in order to quantify the suprathermal electron
     population there.}
  % methods heading (mandatory)
   {A kinetic model for electrons is used which is based on solving the
     Boltzmann-Vlasov equation for electrons including Coulomb collisions with
     both ions and electrons. Initial and chromospheric boundary conditions
     are Maxwellian VDFs with densities and temperatures based on a background
     fluid model. The coronal boundary condition has been adopted from earlier
     studies of suprathermal electron formation in coronal loops.}
  % results heading (mandatory)
   {The model results show the presence of strong suprathermal tails in
     transition region electron VDFs, starting at energies of a few 10
     eV. Above electron energies of 600 eV, electrons can traverse the
     transition region essentially collision-free.} 
  % conclusions heading (optional), leave it empty if necessary 
   {The presence of strong suprathermal tails in transition region electron
     VDFs shows that the assumption of local thermodynamic equilibrium is not
     justified there. This has a significant impact on ionization
     dynamics, as is shown in a companion paper.}

   \keywords{Sun: transition region -- Plasmas -- Methods: numerical}

   \maketitle
%
%________________________________________________________________

%---------------------------------------------------------------------------
\section{Introduction}
%---------------------------------------------------------------------------
Suprathermal tails are a commom feature of solar wind electron velocity
distributions (VDFs). \citet{lin98} has identified different components, a
thermal core, an extended suprathermal halo, and a superhalo with energies up
to tens of keV. \citet{maksimovic97} have shown that kappa distributions
provide better fits to solar wind suprathermal electron VDFs than sums of
Maxwellian distributions. Kappa distributions can also be considered as
equilibrium states in non-extensive statistical mechanics \citep{leubner04,
  livadiotis13}.

While direct observations of suprathermal electron VDFs are only possible
in situ in the solar wind, several models predict a coronal origin. They are,
for example, based on coronal turbulence \citep{roberts98}, obliquely
propagating finite-amplitude electromagnetic waves \citep{vinas00}, or
nanoflares \citep{che14}. In the corona, a suprathermal electron population
can lead to non-local effects through velocity filtration, including heat
fluxes against a temperature gradient \citep{scudder92a, scudder92b}. The
existence of suprathermal electrons corresponds to plasma states away from
thermal equilibrium, whose influence on ion charge states has been discussed
by \citet{cranmer14}, and for the solar wind model of \citet{pierrard14}.

\citet{vocks05} have developed an electron kinetic model for the solar wind
that aims at understanding the formation of a suprathermal electron
component from an initially Maxwellian distribution through resonant
interaction with a given whistler wave spectrum. It is based on solving the
Boltzmann-Vlasov equation, including the effects of Coulomb collisions and
resonant electron -- whistler wave interaction. The wave spectrum has been
chosen as a power law representing the high-frequency tail of wave spectra
associated with coronal heating. The results showed that such a set-up indeed
leads to the formation of suprathermal tails in the electron VDF as a
by-product of coronal heating.
 
\citet{vocks08} applied the same model on the closed volume of a coronal loop
in order to better understand the suprathermal tail formation, without the
open boundary conditions of the outer heliosphere. Indeed, they showed that
power law-like electron VDFs are formed in the keV energy range. This loop
model covers only a spatial domain with relatively high transition region
temperatures near the loop footpoints; the actual transition region was
largely excluded. This was necessary in order to avoid excessive computer
costs associated with small electron thermal speeds and strong spatial density
and temperature gradients there. These simulations confirmed that the quiet
solar corona is capable of producing a substantial suprathermal electron
population.

The hot corona and the cool chromosphere are only separated by a thin
transition region. Coronal suprathermal electrons are not supposed to stop
there. Owing to the $v^4$-dependence of Coulomb collisional mean free paths on
electron speed, $v$, electrons with sufficiently high energy are capable of
traversing the transition region and even entering the cooler and denser
chromosphere.

In this paper, we investigate how such suprathermal electrons propagate from
the corona through the transition region into the chromosphere. The
expectation that suprathermal electrons from the corona can cross the
transition region towards the chromosphere, combined with the strong
temperature gradient of the transition region, leads to the expectation that
transition region electron VDFs can substantially deviate from Maxwellians.

We investigate the evolution of an electron VDF through the strong
temperature gradient of the transition region. The results of this study can
be used as input to calculate ionization states of transition region ions.
The extent and the implications of this influence of suprathermal electrons on
transition region ions, e.g.~concerning the analysis of EUV spectra, will be
discussed in a companion paper.

%---------------------------------------------------------------------------
\section{The model}
%---------------------------------------------------------------------------
The electron kinetic model used here is based on the coronal loop model of
\citet{vocks08}. This is a Boltzmann-Vlasov code, including Coulomb collisions
with both electrons and ions, and resonant interaction with whistler
waves. The assumption of gyrotropic electron VDFs not only reduces the number
of velocity coordinates from 3 to 2, but also eliminates the spatial
coordinates perpendicular to the background magnetic field. Only the spatial
coordinate along a magnetic field line needs to be considered.

The simulation box is now located just below the loop model of
\citet{vocks08}. So the new model is an extension of the old one, with a
simple simulation box that covers the transition region and the uppermost
chromosphere. Owing to its small spatial scale, the magnetic field topology
and its spatial variation do not need to be considered here. The spatial
coordinate of the simulation box is just oriented in the vertical direction,
for a constant magnetic field. Its field strength is the same 136 G as for the
old model loop footpoint, but the actual value has no influence on the results
of the new model.

Resonant interaction with whistler waves is also not considered here; the
``frequency sweeping'' mechanism \citep{tu97, vocks03} would require a
magnetic field variation along the spatial coordinate. Furthermore, this
paper is focused on propagation effects of suprathermal electrons in the
transition region, and the influence of Coulomb collisions on electron VDFs in
cooler and denser solar atmospheric regions. The kinetic model is based on
calculating the temporal evolution of the electron VDF, using the
Boltzmann-Vlasov equation,
\begin{equation}
\frac{\partial f}{\partial t} + (\vec{v} \cdot \nabla) f +
\left[m_e \gamma \vec{g} - e (\vec{E} + \vec{v}\times\vec{B} \right] \cdot
\frac{\partial f}{\partial \vec{p}} = \left(\frac{\delta f}{\delta
  t}\right)_\mathrm{coll},
\label{eq_vlasov}
\end{equation}
until a final steady state has been reached. The parameters $\vec{g}$ and
$\vec{E}$ represent respectively the gravitational and charge separation
electric field, $\vec{B}$ is the background magnetic field, $\gamma = \sqrt{1
+ p^2 / (m_e c)^2}$ is the Lorentz factor, $e$ is the elementary charge, and
$m_\mathrm{e}$ is the electron rest mass. The term on the right-hand side
represents the Coulomb collisions. This Vlasov kinetic electron
model requires a background model for the ions, and also for the electrons as
an initial condition inside the simulation box. A transition region
temperature profile needs to be prepared and hydrostatic equilibrium then
provides a density model. Since there are no heat sources or sinks in the
model, a constant heat flux has to be assumed. The classic Spitzer $T^{5/2}$
law of thermal conductivity in a plasma then leads to the following
temperature profile in a plasma slab,
\begin{equation}
T(s) = \left(T_l^{7/2} + \displaystyle{\frac{s - s_l}{\Delta s}} (T_u^{7/2} -
  T_l^{7/2})\right)^{2/7}
\label{eq_vocks_Ts}
,\end{equation}
with $T_l$ and $T_u$ as lower (chromospheric) and upper (coronal) boundary
temperatures, $s_l$ as the location of the border between the chromosphere and
the transition region, and $\Delta s$ as the transition region thickness. This
leads to a steep temperature gradient at low temperatures. Below the
transition region, the chromospheric temperature profile from Fig.~3 of
\citet{gary07b} has been adopted, which is based on data from
\citet{peter01}. The lower boundary of the simulation box is located in the
uppermost chromosphere, so just the temperature there and the height of the
transition region temperature jump had to be copied. The spatial scale 
of the model transition region from the chromosphere, with $T_l = 1.1\,10^4$
K, to a coronal temperature level, $T_u = 1.4\,10^6$ K, is set to $\Delta s =
500$ km.
\begin{figure}
\includegraphics[width=8.8cm]{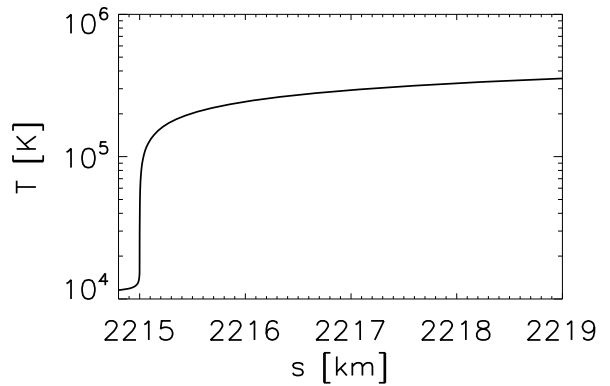}
\caption{Temperature profile of the transition region background model.}
\label{fig_Tprofile_72.eps}
\end{figure}

Figure \ref{fig_Tprofile_72.eps} shows the resulting transition region
temperature profile. An extreme temperature gradient can be seen at lower
temperatures, just above the chromosphere, with temperature changes of the
order of $10^4$ K on just a few meters. According to \citet{lie-svendsen99},
however, the classical heat flux law is still applicable in the 
strong temperature gradient of the transition region, although the presence of
a strong enough suprathermal electron population can lead to deviations from
classical Spitzer conductivity \citep{dorelli99, landi01, dorelli03}.

The initial condition for solving the Boltzmann-Vlasov equation
(\ref{eq_vlasov}) is then provided by Maxwellian VDFs based on the
temperatures and densities of the background model inside the simulation box.
The kinetic model also needs given values for the electron VDF at the
spatial boundaries of the simulation box. For the lower
boundary, this is just a Maxwellian VDF with the density and temperature of
the background model. For the upper boundary, the resulting VDF from the
loop model of \citet{vocks08} with its suprathermal tail is used. The
downward-propagating part of the electron VDF at the footpoint of the loop
model is used as the upper boundary condition for the transition region model
presented here.

\begin{figure}
\includegraphics[width=8.8cm]{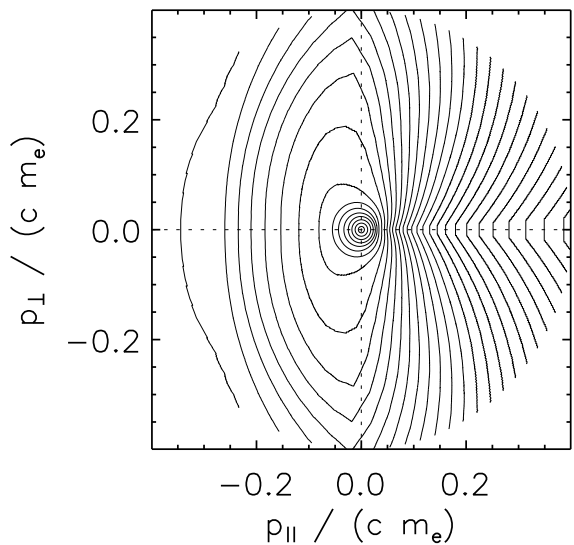}
\caption{Electron VDF at the upper boundary of the simulation box. The
  isolines are chosen in such a way that they form equidistant circles
  for a Maxwellian VDF. A coronal temperature of $T = 1.4\,10^6$ K corresponds
  to a thermal momentum $p_\mathrm{th} = \sqrt{m_\mathrm{e} k_\mathrm{B} T} =
  0.015\,m_\mathrm{e}\,c$.}
\label{fig_vocks_VDF_up}
\end{figure}
This electron VDF is displayed in Fig.~\ref{fig_vocks_VDF_up}. The downward
moving ($p_{\|} < 0$) electron population with a strong suprathermal tail can
be seen. The elongated shape of the VDF is the result of suprathermal electron
production by resonant interaction with whistler waves, as discussed in
\citet{vocks08}. The reason for the strong phase-space gradient near $p_{\|} =
0$ is that in the model of \citet{vocks08}, this VDF was obtained as result
near the loop footpoint where a Maxwellian VDF was provided as a boundary
condition entering the loop ($p_{\|} > 0$). This is not relevant here, since
these electrons are not entering the transition region simulation box.

\begin{figure}
\includegraphics[width=8.8cm]{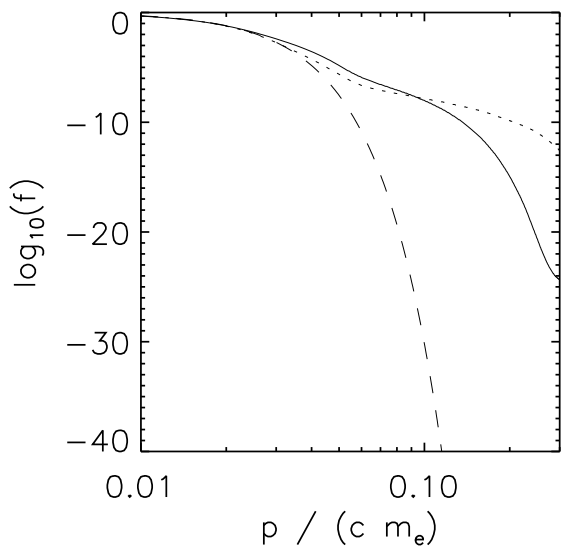}
\caption{Electron VDFs parallel (solid line) and perpendicular (dotted line)
  to the spatial coordinate, and Maxwellian VDF (dashed line) for reference.}
\label{fig_VDF_cut_up}
\end{figure}
Figure \ref{fig_VDF_cut_up} shows cuts through the upper boundary
electron VDF, both parallel and perpendicular to the background magnetic
field, i.e. the spatial coordinate in vertical direction. For the parallel
direction, $p_{\|} < 0$ is plotted. The comparison with a Maxwellian VDF
demonstrates the increase in electron fluxes between $p =
0.05\,m_\mathrm{e}\,c$ and $p = 0.2\,m_\mathrm{e}\,c$, which corresponds to
electron energies of 0.6\,-\,10 keV. These are the suprathermal electrons that
enter the transition region simulation box.

%---------------------------------------------------------------------------
\section{Resulting electron VDF inside the transition region}
%---------------------------------------------------------------------------
After the preparation of initial and boundary conditions, the Boltzmann-Vlasov
equation (\ref{eq_vlasov}) is used to calculate the temporal evolution of the
electron VDF inside the simulation box until a final steady state has been
reached. A simulation time of 0.1 s is sufficient. It allows even
chromospheric thermal electrons to traverse the simulation box multiple
times.

\begin{figure*}
\includegraphics[width=18cm]{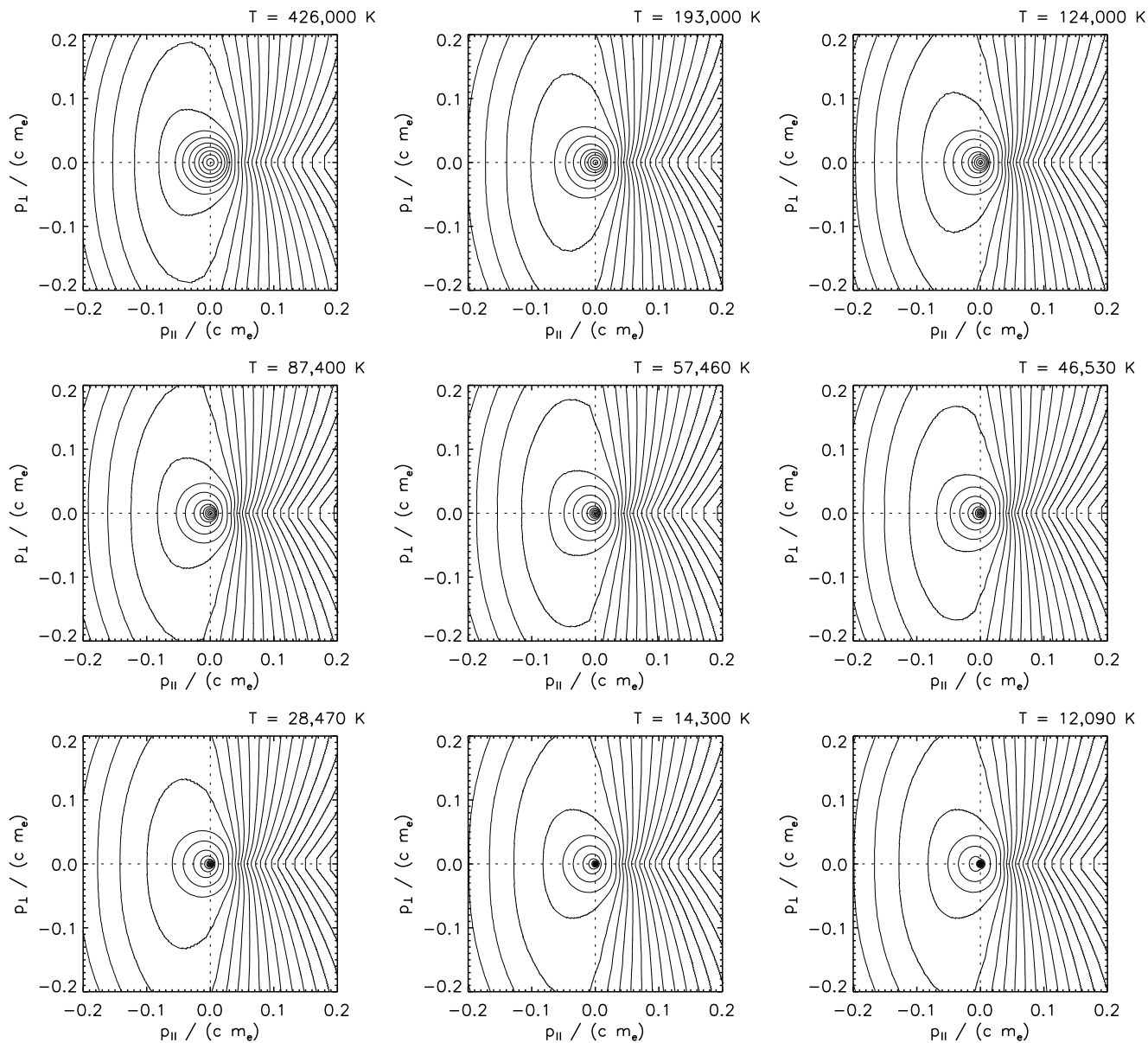}
\caption{Electron VDFs for nine different transition region temperature levels,
  displayed as in Fig.~\ref{fig_vocks_VDF_up}.}
\label{fig_9VDFs}
\end{figure*}

Figure \ref{fig_9VDFs} shows the resulting transition region electron VDFs for
nine different temperature levels. The background temperatures also become
apparent as sizes of the thermal cores of the VDFs. Owing to the
$v^{-3}$-scaling of Coulomb collision frequencies with electron speed, $v$,
the cores are collision-dominated, and the VDFs there quickly approach the
Maxwellians of the background model. On the other hand, the overall shapes of
the VDFs hardly change at higher energies, which is to be expected from the
longer collisional mean free paths there.

\begin{figure}
\includegraphics[width=8.8cm]{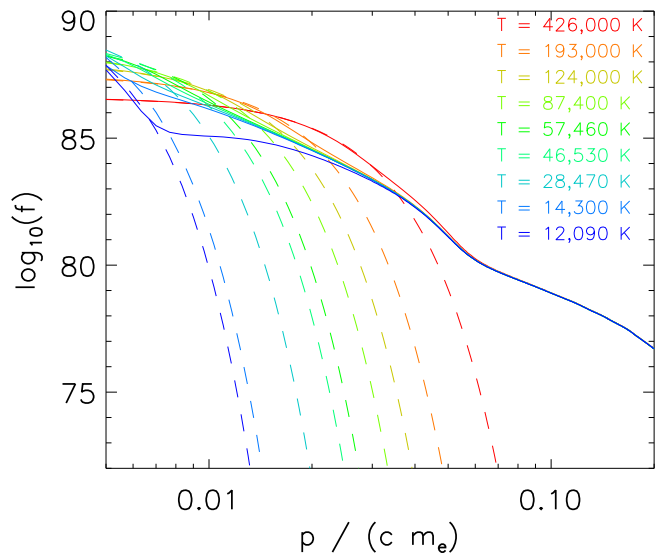}
\caption{Pitch-angle averaged electron VDFs for different transition region
  temperature levels (solid lines), and background Maxwellian VDFs (dashed
  lines).}
\label{fig_eVDFs_72}
\end{figure}

For a better assessment of the suprathermal electron population, the VDFs can
be pitch-angle averaged. Figure \ref{fig_eVDFs_72} shows plots of the
resulting phase space densities for the same nine transition region
temperature levels. Maxwellian VDFs with the same background densities and
temperatures are plotted as dashed lines for reference. For a better comparison
of VDFs with different background densities and temperatures, the plotted
values are not normalized but are shown in absolute units ($\mathrm{kg}^{-3}
\mathrm{m}^{-6}\mathrm{s}^{3}$).

An electron momentum of $p = 0.1\,m_\mathrm{e}\,c$ corresponds to a kinetic
energy of 2.5 keV. The energy range of the suprathermal electrons shown here
therefore covers a few tens of eV up to a few keV.

It can be seen that the phase-space densities are the same for all temperature
levels above an electron momentum of $p = 0.05\,m_\mathrm{e}\,c$, which
corresponds to an energy of 640 eV. These electrons are essentially
collision-free in the transition region, and can traverse it unaffected. For
lower energies, the electron VDFs change with background temperature, as
Coulomb collisions bring them closer to their respective Maxwellians. But even
down to an electron momentum of $p = 0.01\,m_\mathrm{e}\,c$ (25 eV), all lines
are relatively close to each other.

The line for the lowest temperature of 12,000 K, which already corresponds to
the upper chromosphere, is noteworthy. The phase-space density is
significantly lower above $p = 0.01\,m_\mathrm{e}\,c$ (25 eV) than for the next
higher temperature level, which is spatially very close (see
Fig.~\ref{fig_Tprofile_72.eps}). Here, the rapid absorption of low-energy
electrons in the cool and dense chromospheric plasma becomes evident.

\begin{figure}
\includegraphics[width=8.8cm]{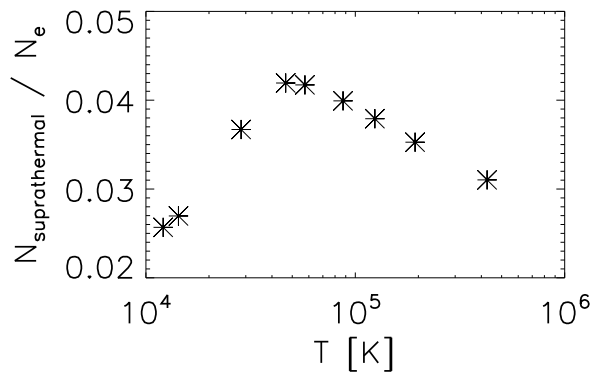}
\caption{Fraction of suprathermal electrons with momentum exceeding $3\times$
  the thermal momentum, as a function of transition region temperature.} 
\label{fig_Nsuprathermal}
\end{figure}

This change in the VDF through the transition region leads to an evolution of
the relative strength of the suprathermal tails as compared to the background
density. The fraction of electrons with momentum greater than 3 times the
thermal momentum, $p > 3 p_\mathrm{th}$, with $p_\mathrm{th} =
\sqrt{m_\mathrm{e} k_\mathrm{B} T}$, is displayed in Figure
\ref{fig_Nsuprathermal}. With decreasing temperature within the transition
region, this fraction first increases. This is due to the small change in the
VDF at suprathermal energies that was noted above, combined with a decreasing
temperature, which lowers $p_\mathrm{th}$, so that a larger fraction of the
phase space is considered ``suprathermal''.

In the lower transition region, however, at temperatures below 50,000 K, this
trend reverses. A local maximum is reached, with 4.2\% of the electrons being
in the suprathermal tail of the VDF. Towards the chromosphere this fraction
decreases rapidly and approaches 2.5\%. This is close to the value of 2.9\%
that corresponds to an isotropic Maxwellian VDF. The slight deviation from
this value is due to the anisotropy of the VDF that is visible in
Fig.~\ref{fig_9VDFs}. This decrease is the consequence of the rapid absorption
of low-energy electrons at energies below 25 eV which can be seen in
Fig.~\ref{fig_eVDFs_72} for the lowest temperature. The remaining suprathermal
tail of the VDF at energies above 25 eV does not contribute significantly to
the fraction $N_\mathrm{suprathermal}$ shown in Fig.~\ref{fig_Nsuprathermal}.

%---------------------------------------------------------------------------
\section{Influence of the transition region background model}
%---------------------------------------------------------------------------
The strong absorption of electrons in the energy range of a few 10 eV
indicates that the thickness of the model transition region itself may have a
strong influence on the resulting suprathermal electron fluxes. In order to
investigate this, we run a second simulation with a slightly modified model
transition region with a flatter temperature profile. It is based on exponents
$5/2$ rather than $7/2$ (Eq. (\ref{eq_vocks_Ts})), as if the plasma thermal
conductivity scales as $T^{3/2}$. This choice is arbitrary, although it
corresponds to the neutral hydrogen thermal conductivity that contributes to
the thermal structure of the chromosphere and lowest transition region at
temperatures $T < 1.5\,10^4$ K \citep{mcclymont83}.
\begin{figure}
\includegraphics[width=8.8cm]{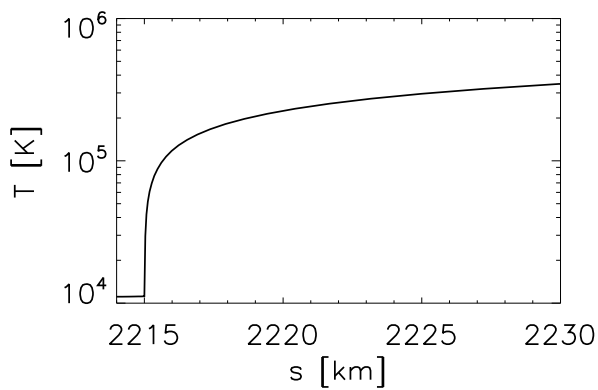}
\caption{Temperature profile of the modified transition region background
  model.}
\label{fig_Tprofile_52.eps}
\end{figure}
The model transition region is now much thicker, see
Fig.~\ref{fig_Tprofile_52.eps}.

\begin{figure}
\includegraphics[width=8.8cm]{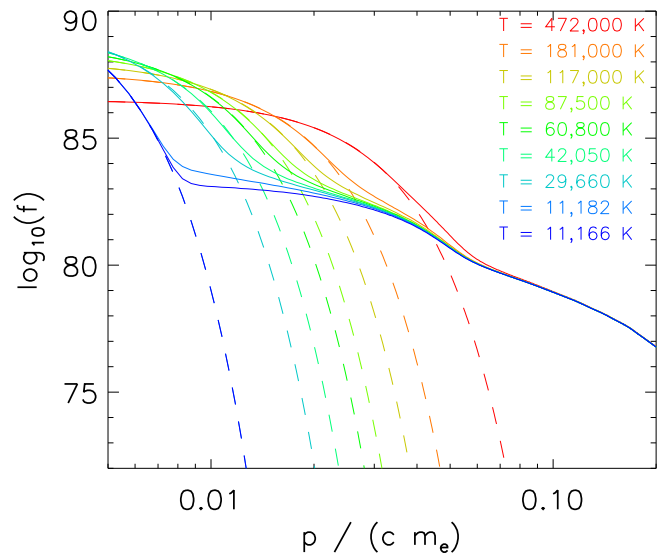}
\caption{Pitch-angle averaged electron VDFs for different transition region
  temperature levels (solid lines), and background Maxwellian VDFs (dashed
  lines), for the modified transition region temperature profile.}
\label{fig_eVDFs_52}
\end{figure}
Figure \ref{fig_eVDFs_52} shows the resulting pitch-angle averaged transition
region electron VDFs for similar temperature levels as in the previous
section. The comparison with the earlier results shows little difference for
higher electron momentum above $p = 0.05\,m_\mathrm{e}\,c$ (640 eV), but for
lower energies the phase-space densities are substantially lower and stay
close to their respective Maxwellian cores. This shows that
the transition region model thickness has a strong influence on suprathermal
electron fluxes in the energy range of a few 100 eV. Because of the
$v^4$-dependence of Coulomb collisional mean free paths, a thicker transition
region allows for the absorption of suprathermal electrons with lower speed.

%---------------------------------------------------------------------------
\section{Discussion and summary}
%---------------------------------------------------------------------------
Our kinetic model results demonstrate that electron VDFs in the transition
region are far away from a local thermodynamic equilibrium state with a
Maxwellian VDF. There are strong fluxes of suprathermal electrons from
the corona which can traverse the transition region essentially
collision-free at energies above 600 eV.

The details of such transition region suprathermal electron fluxes depend on
the coronal electron VDF, but even without a suprathermal electron population
there, the coronal temperature of $1.4\times 10^6$ K can be expected to lead
to suprathermal tails in transition region electron VDFs.

It has been found that the transition region electron phase-space density
surpasses that of a Maxwellian for energies of a few 10 eV and above. This
energy range is crucial for the ionization dynamics and EUV line formation in
the transition region.

The exact values of phase-space densities, and therefore electron fluxes,
depend on the transition region model. A thicker transition region leads to
more suprathermal electron absorption, especially in the energy range up to a
few 100 eV. However, the general result is model-independent: Transition
region electron VDFs show significant suprathermal tails; they are far away
from a Maxwellian except for the collision-dominated thermal core.

This can have significant influence on the analysis of EUV data. A study of
the impact of the suprathermal electron population on transition region
ionization dynamics is presented in a companion paper \citep{dzifcakova16}.

\begin{acknowledgements}
The authors benefited greatly from participation in the International Team 276
funded by the International Space Science Institute (ISSI) in Bern,
Switzerland. E.~D. was supported by the Grant Agency of the Czech Republic
under Grant No. P209/12/1652.
\end{acknowledgements}

% WARNING
%-------------------------------------------------------------------
% Please note that we have included the references to the file aa.dem in
% order to compile it, but we ask you to:
%
% - use BibTeX with the regular commands:
%   \bibliographystyle{aa} % style aa.bst
%   \bibliography{Yourfile} % your references Yourfile.bib
%
% - join the .bib files when you upload your source files
%-------------------------------------------------------------------
\bibliographystyle{aa}
\bibliography{ekinetics_TR.bib}

\end{document}